\documentclass[12pt]{article}
\usepackage[T1]{fontenc}
\usepackage[utf8]{inputenc}
\usepackage{newtxtext,newtxmath}
\usepackage[margin=1.2in]{geometry}
\usepackage[onehalfspacing]{setspace}
\usepackage{booktabs}  %
\usepackage{longtable}  %
\usepackage{graphicx}  %

\usepackage[style=authoryear]{biblatex}
\newcommand{\numVictorianNovelsBassett}{ca. 25,000}  %
\newcommand{\numATCLMarchSnapshot}{15,322}  %
\newcommand{\pageImagesSources}{the Internet Archive, Google Books, HathiTrust, or the British Library}

\newcommand{\numReprintCanonPopulation}{204}
\newcommand{\numReprintCanon}{88}
\newcommand{\numCommonLibraryNovels}{75}
\newcommand{\booksPublishedAnnuallyLateVictorian}{6,000}

\title{Common Library 1.0: A Corpus of Victorian Novels Reflecting the Population in Terms of Publication Year and Author Gender}
\author{
  Allen Riddell \and
  Troy J. Bassett \and
  Laura Schneider\thanks{Equal contributions, order randomized.} \and
  Hannah Mills\footnotemark[1] \and
  Amy Yarnell\footnotemark[1] \and
  Rachel Condon\thanks{Equal contributions, order randomized.} \and
  Joseph Bassett\footnotemark[2] \and
  Sara Duke\footnotemark[2]
}
\date{2019-08-29}

\begin{document}

\maketitle

\begin{abstract}

  Research in 19th-century book history, sociology of literature, and quantitative literary history is blocked by the absence of a collection of novels which captures the diversity of literary production.
  We introduce a corpus of \numCommonLibraryNovels{} Victorian novels sampled from a \numATCLMarchSnapshot{}-record bibliography of novels published between 1837 and 1901 in the British Isles.
  This corpus, the Common Library, is distinctive in the following way: the
  shares of novels in the corpus associated with sociologically important subgroups match the shares in the broader population.
  For example, the proportion of novels written by women in 1880s in the corpus is approximately the same as in the population.
  Although we do not, in this particular paper, claim that the corpus is a representative sample in the familiar sense---a sample is representative if
``characteristics of interest in the population can be estimated from the sample with a
known degree of accuracy'' \autocite[3]{lohr2010sampling}---we are confident that the corpus will be useful to researchers.
  This is because existing corpora---frequently convenience samples---are conspicuously misaligned with the population of published novels.
  They tend to over-represent novels published in specific periods and novels by men.
  The Common Library may be used alongside or in place of these non-representative convenience corpora.

\end{abstract}

Those studying and teaching the Victorian novel routinely work with
non-representative corpora of novels. The most accessible novels from the period
are those which are still in print—for example, the collection of
\numReprintCanonPopulation{} novels still in print in popular series such as Oxford World Classics and Penguin Classics. However, these popular reprint series have drawbacks since they over-represent novels written by
men and novels first published in the 1860s
\autocites{bassett2017median}{riddell2018reassembling}. The absence of a
representative corpus is a particular problem for those interested in the
social history of literature and those studying the production
and spread of literary forms across national and linguistic borders \autocites{williams1961social, moretti2000conjectures, bode2018world}.
Without a representative corpus it is difficult to analyze, say, the relationship (if any)
between writers' socio-economic background and the content or style of
their writing. It is also difficult to detect the emergence or spread of literary techniques over time and space.

The absence of a representative corpus is understandable since there is no exhaustive list
of Victorian novels on the basis of which one might construct a corpus via
simple random sampling. That no such list exists is due to the large number of books
published during the period and the considerable labor involved in identifying novels.
Identifying a novel (vs. a non-novel) often requires inspecting a physical
copy (or digital surrogate) of a book since titles alone often mislead or fail to indicate the genre—e.g., Charlotte Brontë's first novel had the title ``Jane Eyre: An Autobiography''.
The number of books which need to be inspected
by a domain expert is large: during the last decade of the 19th century publishers in the British Isles
issued roughly \booksPublishedAnnuallyLateVictorian{} books (novels and non-novels) each year \autocites[294]{eliot2012few}[123]{eliot1994patterns}. Although researchers
expressed interest in a better accounting of careers of novels and novelists during the
period, the resources required for an exhaustive bibliography were never marshalled\autocite[588-589]{sutherland1988publishing}.

The corpus accompanying this paper---Common Library version 1.0---supports research in (quantitative) literary history and corpus linguistics. (Table~\ref{tbl:novels-list} lists the titles in the corpus.).
Unlike other available (convenience) samples of 19th-century novels, the proportions of novels in the Common Library associated with each year between 1837 and 1901 mirror the proportions in the target population.
For each year the proportions of novels associated with authors of different genders (men, women, and unknown) also reflects estimated shares in the population (Table~\ref{tbl:novels-counts-by-year-and-gender}). For example, 27\% of novels in the Common Library are novels by women written between 1876 and 1901 (inclusive). This percentage matches the corresponding percentage (28\%) in the population. Deviations from the population distribution are due to natural variability in random sampling.

\begin{table}
\centering
\small
\begin{tabular}{lllllll}
\toprule
period & \multicolumn{3}{l}{1837-1875} & \multicolumn{3}{l}{1876-1901} \\
gender &           m &           w &          u &           m &           w &          u \\
label              &             &             &            &             &             &            \\
\midrule
Common Library 1.0 &    18 (24\%) &     9 (12\%) &     3 (4\%) &    23 (31\%) &    20 (27\%) &     2 (3\%) \\
Reprint Canon      &    70 (34\%) &    32 (16\%) &     0 (0\%) &    86 (42\%) &     16 (8\%) &     0 (0\%) \\
Population (est.)  &  3754 (15\%) &  3715 (15\%) &  1256 (5\%) &  8124 (32\%) &  6900 (28\%) &  1323 (5\%) \\
\bottomrule
\end{tabular}

\caption{Counts of novels in the Common Library 1.0, the Reprint Canon, and the broader population of novels published in the British Isles, 1837--1901.}\label{tbl:novels-counts-by-year-and-gender}
\end{table}

\section{Previous Research and Related Data}

Population growth, new technologies, and new financial institutions contributed
to exponential growth in the rate at which previously unpublished novels (``new
novels'') appeared between 1837 and 1901 in the British Isles. At the start of
this period roughly 100 new novels appeared each year.  By the end of the
period, publishers produced over 1,000 new novels every year. In a period
during which the population roughly doubled, we witness a tenfold increase in
the rate of production of new novels and, as the size of print runs did not decrease, in novel copies generally
\autocite[294]{eliot2012few}.\footnote{According to \textcite{eliot2012few},
14,550 book titles (novels and non-novels) appeared in the decade starting with 1800 and 60,812
appeared during the 1890s. Print runs of the most successful books increased by
a factor of roughly eight during the period. Given this, a ten fold increase in
the number of book copies produced in 1837 versus 1901 seems possible.} Much
of this increase is likely due to the declining cost of paper and declining
costs associated with printing.  Steam-powered papermaking and steam-powered
printing were widely adopted by mid-century
\autocites[64]{weedon2003victorian}[224]{raven2007business}. Financial
institutions also matured, especially during the 1830s. More mature financial
institutions further lowered costs to publishers as they made raising money to
pay for capital improvements and new publications less expensive
\autocite[62]{weedon2003victorian}. An expanding
population of readers able to afford access to novels also likely contributed to the growth in new novel
production. One number makes the scale of the expansion clear: the median
publication year for a Victorian novel is in the mid 1880s. As many new titles appeared
during the 16 years after 1884 as appeared during the preceding 47 years.

Literary historians, book historians, and sociologists of print culture have aspired to a
fuller view of the population of \numVictorianNovelsBassett{} novels---their particular
morphology, style, syntax, etc.---as well as of the novelists involved---their lives,
social background, professional networks, etc. The scope for learning here, or,
alternatively, the extent of literary historians' ignorance is immense. John
Sutherland, the doyen of Victorian literary history, laments the ``sheer unavailability of
necessary empirical knowledge'' for research, adding that ``one of the things that makes
literary sociology so easy to do at the moment is that we don't know enough to make it
difficult'' \autocite[558]{sutherland1988publishing}.  Literary historians do not know,
for example, how many people pursued careers as novelists in the British Isles during the
19th century
\autocites[574-575]{sutherland1988publishing}[151-164]{sutherland1995victorian}.
This ignorance persists despite decades of sustained scholarly activity on the Victorian
novel at research universities across the world.

Many researchers find the present situation unsettling, particularly in light
of the standard classroom practice of teaching the history of the Victorian
novel using a small number of novels by canonical authors
\autocites[207-210]{moretti2000slaughterhouse}[27]{bode2018world}[87]{bode2017equivalence}.
This approach neglects discussion of the broader literary system. In
particular, it ignores the range and variety of other novels which readers would have
encountered at the same time as they encountered a canonical novel. The
standard presentation of the Victorian novel also typically neglects discussion
of the range of material and economic forces or the range of intermediaries
---e.g., booksellers, reviewers, advertisers, circulating libraries, book clubs---operating in the literary market. Serious discussion of these would require
more organized information about the period than literary historians currently have
available.

Research of practical value to those interested in a more extensive and
democratic account of the novel has tended to come, especially in recent
decades, not from literary studies but rather from scholarship allied with book
and publishing history. Work with a machine-readable version of the
\emph{Nineteenth-Century Short Title Catalog} (NSTC) in
\textcite{eliot1994patterns} and \textcite{eliot1997patterns} is a notable
example. \textcite{eliot1997patterns} delivers a time series which describes
the total number of books published in London, Oxford, Cambridge, Edinburgh,
and Dublin each year between 1801 and 1870. Combined with weekly records of
publisher-reported editions in \emph{Publishers' Circular} it is possible,
as \textcite{weedon2003victorian} shows, to estimate the annual number of books (novels
and non-novels)
published in the British Isles. Although this series does not directly tell us
the number of (new) novels published during the period, it does bound from above the
annual number of novels published.  And if we are willing to assume the
percentage of books which are novels does not change radically from year to year during
the 19th century, the time series tells us a great deal about how the rate of novel publication
changes over the period.

If the work of Eliot and others gives us a sense of the maximum number of previously unpublished
novels appearing each year, \textcite{bassett2018atcl} provides confidence in the minimum number of new novels published each year.
Started in 2007, the online database \emph{At the Circulating Library} (ATCL) contains records of
more than 17,000 titles published between 1837 and 1901. (In this paper we use a March
2018 snapshot of the database with \numATCLMarchSnapshot{} titles.) Although ATCL aspires to provide an
exhaustive list of new novels, it is, as of this writing, incomplete. ATCL's coverage is not
uniform, some years are virtually complete (e.g., 1838) whereas others (e.g., 1900) are perhaps only 60\%
complete. Even in the incomplete years, ATCL bounds from below the number of new novels. The
database tells us the minimum number of new novels which appeared in each year.

Another stream of research, intermittently connected with book and publishing history, has
used convenience sampling of digitized novels in work which purports to provide
information about the population of novels published during the 19th century
\autocites{jockers2013macroanalysis,jockers2013significant,
algee-hewitt2016canon,underwood2018transformation,piper2018enumerations}. For example,
\textcite{jockers2013macroanalysis} uses a corpus of 3,346 novels and
\textcite{algee-hewitt2016canon} uses a corpus of 1,117 novels. All these convenience corpora are gathered opportunistically from texts available
from commercial providers or from digitizations of works found in North American and UK
libraries. In \textcite{jockers2013macroanalysis} and \textcite{algee-hewitt2016canon},
the researchers admit that their corpora are, in fact, non-probability convenience samples
\autocites[172]{jockers2013macroanalysis}[2-3]{algee-hewitt2016canon}.

\section{Methods}

The novels which are included in the Common Library are gathered using a probability-proportional-to-size sampling strategy.
First, a publication year and author gender pair is sampled according to its share in the population.
Previously published estimates of the number of new novels published by year and author gender provide the information we need for this first step \autocite{riddell2018reassembling}.
Then a novel with a matching publication year and author gender is sampled uniformly at random from the \emph{At the Circulating Library} (ATCL) database.
If the novel does not have a publicly-available digital surrogate, the sampling is repeated until one with a surrogate is found.
Having selected a specific novel, we then randomly sample a chapter and key-in the chapter text. These chapter texts comprise the Common Library.
This sampling strategy does not, needless to say, yield a simple random sample from the population. It does not yield a representative sample either. (A sample is representative if
``characteristics of interest in the population can be estimated from the sample with a
known degree of accuracy'' \autocite[3]{lohr2010sampling}.) This is because the sampling frame is restricted to novels with digital surrogates and the ATCL is not, at this point, exhaustive.
The strategy does, however, ensure that, with respect to publication year and author gender, shares of novels in the Common Library are aligned with corresponding shares in the population.

The following example illustrates the two-stage sampling procedure.
Of the \numVictorianNovelsBassett{} novels published during the period, 288 (1.1\%) were published in 1894 and written by women authors.
The probability of sampling this group in the first stage is 1.1\%.
Assuming we sampled this group, we would then sample a specific novel at random from the list of
novels in the ATCL database which have matching year and author gender until one is
located which has a first edition digital surrogate. With a specific novel in hand, we
randomly sample a chapter and manually key-in its text.

A novel associated with this particular group (1894, woman-authored) appears in the Common Library.
The group was sampled on the 21st probability-proportional-to-size draw and the following novel was sampled from ATCL:
Edna Lyall's \emph{To Right the Wrong} published in 1894 by Hurst and Blackett.
This three-volume novel has 41 chapters and the 25th chapter was randomly sampled for encoding.
For reproducibility, we use the novel's ATCL database identifier as a random seed when sampling the chapter.\footnote{For example, using Python 2, {\tt random.seed(4705); random.randrange(41) + 1} yields 25 (for chapter 25). The addition of 1 to the sampled integer is required because {\tt random.randrange(41)} samples an integer uniformly at random from the interval $[0,41)$. Since the chapters we are interested in are numbered from 1 to 41 (inclusive), we add 1 to the result.}

This section is organized as follows. First we describe in detail the two-stage
sampling procedure: (1) the partitioning of the population into subpopulations
defined by publication year and author gender and (2) the use of the ATCL
database to select a novel for each sampled group.

\paragraph{New novels by year and author gender, 1837-1901.} We begin by sampling a group defined
by publication year and author gender. As each novel has a distinct publication year and author gender, these groups partition the population.
We sample a group with probability proportional to the number of novels associated with it.
For the sizes of the groups we use medians of the estimated counts in \textcite{riddell2018reassembling} (Table~\ref{tbl:gender-year-cluster-sizes}).
Publication year is the year indicated on the title page of the first edition.
Author gender is the gender of the historical individual acknowledged as the novel's author.
For pseudonymous and anonymous novels which advertise an author gender on the title page (e.g., ``By a Lady of Rank''), we use the advertised gender. Here we
follow the convention established by \textcite{garside2004british} in assuming that the gender of the historical individual(s) who wrote the novel is the same as the gender of the advertised author.
When no information is available from the title page about an anonymous or pseudonymous author, we record the author's gender as ``unknown''.
In the exceedingly rare case of a title with more than one author, we use the gender of the author listed first on the title page of the first edition.

One technical detail concerning the definition of ``novel'' deserves to be mentioned.
There are two definitions of the novel used in large bibliographies of the 19th-century English novel:
the descriptive definition found in \textcite{raven2000english} and \textcite{garside2000english} (hereafter ``RFGS'') and the definition used by ATCL (hereafter, ``ATCL'').
A novel according to RFGS is a book described as a ``novel'' by contemporaries.
The definition used by ATCL is more permissive:
prose fiction of at least 90 printed pages that is not addressed exclusively to children.
These definitions are largely consonant.
All books which are novels according to RFGS are novels according to ATCL.
Some books considered novels by ATCL are not novels according to the more restrictive RFGS definition.
Disagreements are predictable as they tend to concern novel-like books in well-known subgenres.
ATCL's definition permits novel-like works of (didactic) religious and juvenile fiction to be counted as novels;
RFGS exclude these books.
We discuss these two definitions and list conforming examples in Appendix~\ref{appendix:novel-definitions}.
In the sampling strategy used here, we make the assumption that during the 1837-1901 period any count
of novels using the ATCL definition is equivalent to a count using the RFGS definition after increasing the latter by 12.5\%.
This figure is the midpoint of the estimate that between 10\% and 15\% of ATCL titles would be excluded from a RFGS bibliography, were RFGS to cover the 1837-1901 period.
Future work might revisit this 12.5\% figure if, say, men are much more likely to be authors of works which would be excluded by the RFGS standard (e.g., novel-like didactic religious fiction).
Although \textcite{riddell2018reassembling} use the RFGS definition of the novel, the sampling probabilities do not change after translating the counts into ATCL terms.

\paragraph{\emph{At the Circulating Library} Bibliography, \numATCLMarchSnapshot{} titles.}
Once we have sampled a group of novels defined by publication year and author gender, we
sample a title uniformly at random from titles with matching publication year and author
gender in the \emph{At the Circulating Library} (ATCL) database.\footnote{
  For those interested in reproducing the sample, we provide the random seed and software version used.
  The SQL command which performs the sampling is the following: \texttt{SELECT author\_id, title\_id FROM titles LEFT JOIN authors USING (author\_id) WHERE publication\_year=? AND gender="?" ORDER BY RAND(1);}.
  MySQL version 5.7.21 was used.
}
We use the March 7, 2018
snapshot of the database. This version includes \numATCLMarchSnapshot{} titles published between
1837 and 1901. If the title sampled has no publicly available first-edition page images, we sample another title until we find one which does.
In order for a title to be
counted as having a first-edition digital surrogate, page images of the first edition must
have been available online on or before December 31, 2018 from \pageImagesSources{}.
We count other editions published by the first-edition publisher \emph{in the same year} as the first edition as first editions.
These editions include second printings (sometimes labeled ``second edition'') as well as export editions of novels, which often feature a variant title page.\footnote{
  For example, we count a Canadian edition of Swan's \emph{Mrs.  Keith Hamilton} (1897)
  as a first edition as the book was printed by the London publisher
  in the same year as the London edition and appears to be identical, save for an alternate
  title page.}
We count a multivolume novel as having a first-edition digital surrogate if
all volumes of the first edition have been digitized.  For example, the novel mentioned
earlier---Lyall's \emph{To Right the Wrong}---is a three-volume novel (``triple decker'')
with first-edition digital surrogates of all three volumes. (The volumes were digitized in
October 2008 by the Internet Archive from originals at the University of Illinois
Urbana-Champaign.)

Once we have a title with an available first-edition digital surrogate, we sample a
chapter uniformly at random and then manually key-in the sampled chapter.  We encode a single
chapter to limit the time and resources expended on keying-in texts. We welcome reports of
discrepancies between observed characteristics in the chapters encoded and page images of
the first edition. If the sampled novel is not divided into chapters, we sample a section
at random. (Every novel sampled was explicitly divided into chapter-like sections of some kind.) Non-narrative sections of the text (e.g., backmatter, dedications, non-narrative authors' prefaces) are not counted as sections. We
further annotate the UTF8-encoded ``plain text'' version of each chapter using a small set
of HTML5 tags. For example, we mark italics and paragraph boundaries.  We label the resulting
collection of 75 novel chapters as the ``Common Library''. The version of the collection described here is 1.0. The dataset accompanies this paper.\footnote{The filename is {\tt common-library-v1.0.zip}. The file contents have the following SHA256 hash: {\tt caa78440bbc237b1bfa3600dd3b2577263fed3cbde723dbc2adf250a0fae6b58}.}

\paragraph{Over-representation of multivolume novels and multiple-novel author novels in
\emph{At the Circulating Library}.} The ATCL database aims to be exhaustive but the March
7, 2018 snapshot we use does not contain a record for every published novel during the
Victorian period. Although the database has grown primarily by the systematic addition of
titles from annual lists of titles published in \emph{The English Catalogue of Books},
titles also arrive in the bibliography via other routes, resulting in over-representation
of certain kinds of novels: multivolume novels and novels by authors who published more than one novel. For example,
whereas about 29\% of Victorian novels are multivolume novels, 57\% of the Common Library titles are multivolume novels.
In a future article we will quantify the biases precisely.

\paragraph{Restriction to first-edition digital surrogates.} In addition to being
restricted to ATCL titles, our sampling frame is further restricted to titles with
first-edition digital surrogates created on or before December 31, 2018. This biases the
novels in the Common Library towards novels which were targets for library collection
during the 19th and 20th century. The reason for this is that first editions which survive
in a greater number of libraries are more likely to be found at a library which digitized
its holdings.

We do not know precisely what made novels targets for library collections. Nearly all novels from
the period survive in libraries. For example, 98\% of titles published in 1838 have a
surviving copy. Although only some libraries participated in library digitization, two of
the libraries which did, Oxford's Bodleian and the British Library, tended to receive a
copy of every novel published as a consequence of the library deposit requirement
introduced by the Copyright Act of 1842. (The Copyright Act's deposit requirement
specified the British Museum, whose library department is now the British Library.) The
library deposit requirement should allay concerns that Oxford or the British Library
tended to exclude certain kinds of novels, at least after 1842.

Concern about the first-edition digital surrogate limitation introducing
unaccounted-for bias into the sample is warranted. Oxford's Bodleian did not digitize its
entire collection and parts of the British Library's collection were destroyed during
World War II. So novels which were targets of collection by other libraries are more
likely to have first-edition digital surrogates. This means that novels which were targets
for collection are more likely to be in the Common Library than novels which were not
targets.\footnote{Future work might approach this problem by treating books which have
first-edition digital surrogates from the Bodleian or British Library---where legal
deposit can be assumed to be the reason the book is present---differently than books which
are not available from those two libraries.}

For novels associated with a given year and author gender, we believe a promising account of which
novels were targets for collection is available. We hypothesize that novels written by novelists who wrote at least one other novel---multiple-novel author titles---are more likely to be targets for collection than novels by single-novel authors. We make use of ATCL's exhaustive coverage of novels published in 1838 to evaluate this hypothesis. Comparing 1838 novels by authors of the same gender
with first-edition digital surrogates to novels without surrogates provides preliminary, non-decisive evidence in support of the hypothesis. Whereas 6 of 9 novels (66\%) published in 1838 by single-novel women novelists have first-edition
digital surrogates, 24 of 28 novels (86\%) by multiple-novel women novelists have first-edition
digital surrogates. We suspect that authors of a commercially or
critically successful novel was encouraged to write---as they are today---additional
novels by bookseller-publishers and other text industry intermediaries. Commercially and
critically successful novels were, we believe, targets for collection. These novels would be more likely to be requested by patrons and many libraries do respond to patrons' demands. With respect to each subpopulation defined by a given publication year and author gender, restricting the Common Library
collection to novels with first-edition digital surrogates likely increases the tendency for the Common Library to over-represent multiple-novel authors.

\section{Common Library vs. Reprint Canon}

Our primary contribution is the \numCommonLibraryNovels{}-novel Common Library
corpus, a sample from the population of Victorian
novels. With respect to groups defined by publication year and author gender,
the shares of novels in the Common Library associated with each group reflect
the shares of novels associated with the group in the population.

In this section we offer a superficial comparison of the Common Library with a familiar collection of Victorian novels. This other corpus, which we label the ``Reprint Canon'', consists of the 204 Victorian novels which are in print and available from Penguin, Oxford, or Broadview in 2017.\footnote{ We use the 2017 catalogs from Penguin
  (Classics), Oxford (World Classics), and Broadview.  Although the set of titles in print does change occasionally, there is typically little or no change from year to year.
} We study this corpus through the Reprint Canon Sample, a simple random sample from the Reprint Canon. The Reprint Canon Sample contains \numReprintCanon{} novels.
We encode one chapter, chosen at random, from each Reprint Canon novel using the same procedure we used with the Common Library.
We compare the Common Library to the Reprint Canon Sample using a simple feature, length in words.
We approximate length by multiplying the number of chapters in each novel by the length in words of the encoded chapter.

\begin{figure}
    \centering
    \includegraphics{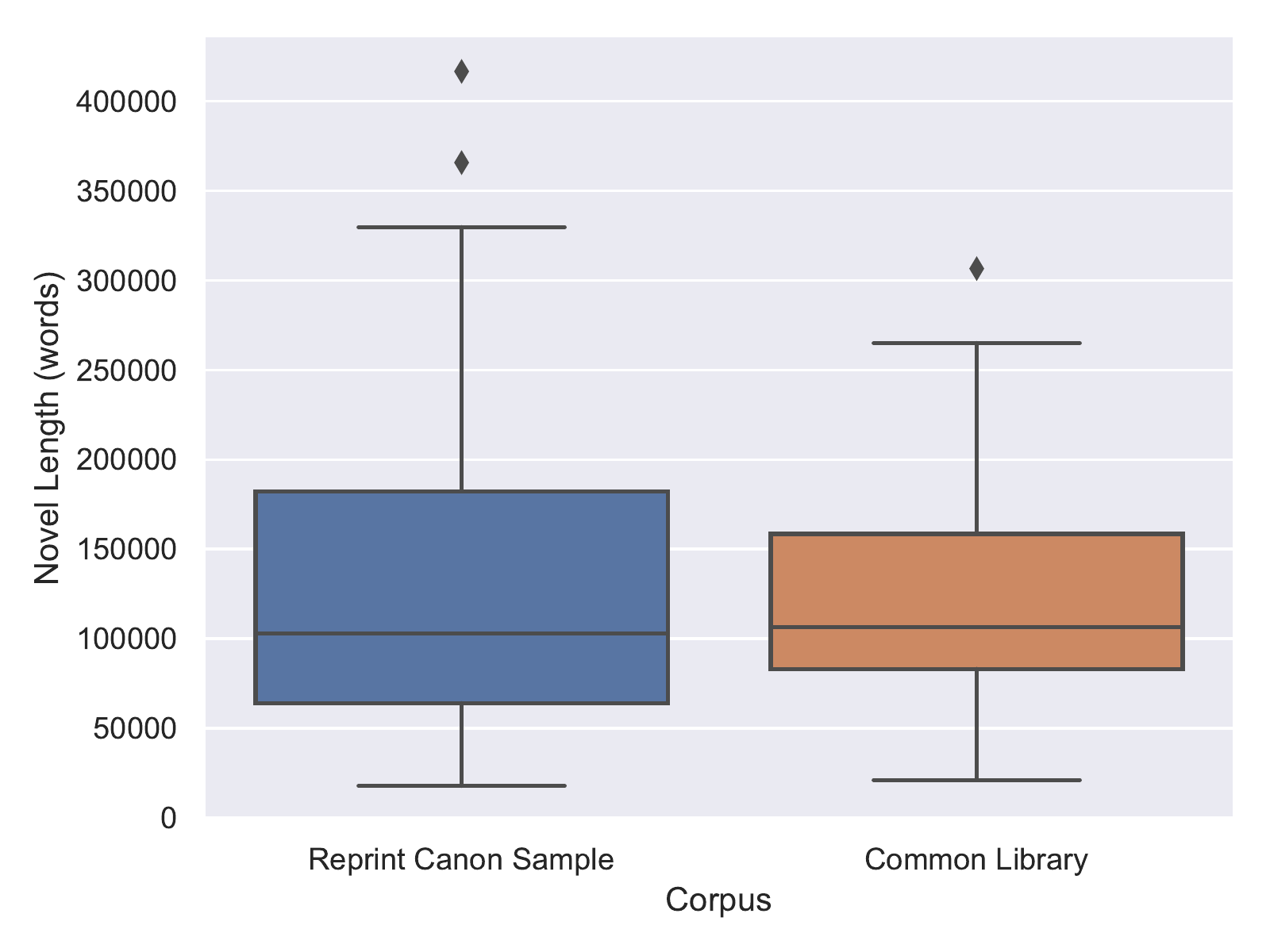}
    \caption{Estimated Novel Lengths in the Common Library ($n=75$) and the Reprint Canon Sample ($n=88$). Novel length calculated by multiplying the length (in words) of the encoded chapter by the number of chapters.}
    \label{fig:features-length}
\end{figure}

The Common Library novels differ from the Reprint Canon Sample in terms of length.
The mean length in the Common Library collection is conspicuously smaller than the mean length of Reprint Canon Sample novels (Figure~\ref{fig:features-length}).
The reason for this is the prevalence of very long novels in the Reprint Canon Sample. Very long novels (> 250,000 words) are more common in the Reprint Canon Sample than in the Common Library. The Reprint Canon Sample has 12 (14\%) very long titles. The Common Library, by contrast, has only 3 (4\%).

This difference is suggestive, provided we accept that the Common Library novels do better reflect the diversity of novels published during the Victorian period.
The difference should call into question the claim that the novels in the Reprint Canon capture the diversity of prose fiction published between 1837 and 1901.
This claim is explicitly or implicitly present in university courses on the Victorian novel.
It remains standard
classroom practice to teach the history of the Victorian novel using only novels which are part of the Reprint Canon \autocites[207-210]{moretti2000slaughterhouse}[27]{bode2018world}[87]{bode2017equivalence}.
Although this approach is regarded as neglecting discussion of the broader literary system, it has been difficult to find specific evidence that backs up such a belief.
The difference observed here---combined with the assumption that the Common Library novels do better reflect the diversity in the population---supplies this evidence.

\section{Conclusion}

This paper makes available a collection of \numCommonLibraryNovels{} novels from the Victorian period which reflects the larger population of novels in terms of publication year and author gender.
We label this corpus the Common Library.
Saying that the collection reflects the population in terms of publication year and author gender means that, for example, the share of novels in the collection published in the year 1888 by women writers matches the share of such novels found in the population.
In other respects, the collection reflects the population poorly.
Multivolume novels and novels by authors of more than one novel are over-represented.
(In a future publication we will describe how these biases can be adjusted for using post-stratification.)
Other publicly available collections of novels from the period plainly do not reflect the population of published novels.
One prominent and widely-used collection, Victorian novels in print today (the 204-novel Reprint Canon), over-represent novels by men writers and novels published during the 1860s.
The Common Library, by contrast, better reflects the diversity of novels published during the Victorian period.

\section*{Contributions}

AR and TJB planned the research and wrote the paper. All individuals contributed encoded texts. AY and AR organized and reviewed the encoded texts.

\printbibliography

\pagebreak

\section*{Appendix: List of Novels in the Common Library Collection}

\scriptsize
\begin{longtable}{lllrlll}
\caption{Novels in the Common Library 1.0 corpus. Note: ``BTAO'' stands for ``By the Author of.''}\label{tbl:novels-list} \\
\toprule
{} &            Last Name &                                    Title &  Year & Gender & Single-Novel Author & Multivolume \\
ATCL ID &                      &                                          &       &        &                     &             \\
\midrule
\endhead
\midrule
\multicolumn{7}{r}{{Continued on next page}} \\
\midrule
\endfoot

\bottomrule
\endlastfoot
15732   &               Nisbet &        Comrades of the Black Cross: A... &  1899 &      M &                  no &          no \\
1237    &               Curtis &                     Favourite of Fortune &  1886 &      W &                  no &         yes \\
7083    &                Gleig &                               The Hussar &  1837 &      M &                  no &         yes \\
4847    &           Churchward &           Jem Peterkin's Daughter: An... &  1892 &      M &                  no &         yes \\
11712   &               Brooks &   The Naggletons, and Miss Violet and... &  1875 &      M &                  no &          no \\
9982    &                Adams &           The White Brunswickers: or,... &  1865 &      M &                  no &          no \\
951     &               Gibbon &           The Braes of Yarrow: A Romance &  1881 &      M &                  no &         yes \\
15354   &                Frith &    In Search of Quiet: A Country Journal &  1895 &      M &                  no &          no \\
14740   &              Prevost &                            Entanglements &  1898 &      M &                  no &          no \\
7173    &              Cummins &                      El Fureidis: A Tale &  1860 &      W &                  no &         yes \\
8550    &              Boothby &                      Pharos the Egyptian &  1899 &      M &                  no &          no \\
1053    &            Alexander &                       The Admiral's Ward &  1883 &      W &                  no &          no \\
1519    &                 Lyon &                     The Signora: A Novel &  1883 &      M &                  no &         yes \\
10236   &                Green &  The Heir of Hascombe Hall: A Tale of... &  1900 &      W &                  no &          no \\
7155    &            Thackeray &      Vanity Fair: A Novel without a Hero &  1848 &      M &                  no &          no \\
1120    &                Booth &                      Fragoletta: A Novel &  1881 &      W &                  no &         yes \\
11919   &             Clifford &  In a Corner of Asia: Being Tales and... &  1899 &      M &                  no &          no \\
502     &               Murray &                John Alston's Vow: A Tale &  1865 &      W &                  no &         yes \\
2534    &               Knight &    A Romance of Acadia Two Centuries Ago &  1874 &      M &                  no &         yes \\
1453    &              Jenkins &                         Jobson's Enemies &  1882 &      M &                  no &         yes \\
4705    &                Lyall &                       To Right the Wrong &  1894 &      W &                  no &         yes \\
6459    &         Peregrinator &     A Flight to Florida: And All That... &  1888 &      U &                 yes &         yes \\
7900    &              Terrell &                     The City of the Just &  1892 &      M &                  no &          no \\
12419   &             Marshall &       Constantia Carew: An Autobiography &  1883 &      W &                  no &          no \\
7741    &            Alexander &                  Through Fire to Fortune &  1900 &      W &                  no &          no \\
12089   &               Croker &            Angel: A Sketch in Indian Ink &  1901 &      W &                  no &          no \\
14640   &                 Hood &  A Disputed Inheritance: The Story of... &  1863 &      M &                  no &          no \\
3583    &                James &       The Fate: A Tale of Stirring Times &  1851 &      M &                  no &         yes \\
2360    &                 Hill &            They Were Neighbours: A Novel &  1878 &      M &                 yes &         yes \\
13760   &              Conyers &                            The Thorn Bit &  1900 &      W &                  no &          no \\
552     &                Platt &                             Angelo Lyons &  1866 &      M &                  no &         yes \\
8440    &              Raymond &                        Two Men o' Mendip &  1899 &      M &                  no &          no \\
956     &               Gibbon &                         The Golden Shaft &  1882 &      M &                  no &         yes \\
7143    &                Power &                              Nelly Carew &  1859 &      W &                  no &         yes \\
1523    &              MacEwen &             Miss Beauchamp: A Philistine &  1883 &      W &                  no &         yes \\
4783    &               Sherer &   Alice of the Inn: A Tale of the Old... &  1893 &      M &                  no &         yes \\
1012    &              Marryat &              Peeress and Player: A Novel &  1883 &      W &                  no &         yes \\
8165    &                 Swan &       Mrs. Keith Hamilton, M.B.: More... &  1897 &      W &                  no &          no \\
4228    &                Peake &  Cartouche, the Celebrated French Robber &  1844 &      M &                 yes &         yes \\
1359    &                 Gray &                 The Reproach of Annesley &  1889 &      W &                  no &         yes \\
15770   &             Chetwode &           The Knight of the Golden Chain &  1898 &      U &                  no &          no \\
1289    &                 Fane &              The Story of Helen Devenant &  1889 &      W &                  no &         yes \\
4651    &              Gissing &                 The Emancipated: A Novel &  1890 &      M &                  no &         yes \\
6565    &            Coleridge &                                Waynflete &  1893 &      W &                  no &         yes \\
424     &              Harwood &                   Lord Ulswater: A Novel &  1867 &      M &                  no &         yes \\
8557    &              Boothby &                       Long Live the King &  1900 &      M &                  no &          no \\
4040    &  BTAO Sin and Sorrow &   Sin and Sorrow: A Story of a Man of... &  1850 &      U &                  no &         yes \\
4844    &               Castle &                    Consequences: A Novel &  1891 &      M &                  no &         yes \\
6797    &              Marston &       James Vraille: The Story of a Life &  1890 &      M &                 yes &         yes \\
2623    &                 Payn &             A Woman's Vengeance: A Novel &  1872 &      M &                  no &         yes \\
4383    &           Henningsen &      The White Slave: or, The Russian... &  1845 &      M &                  no &         yes \\
1256    &                Diehl &                      Eve Lester: A Novel &  1882 &      W &                  no &         yes \\
7951    &                James &                    Tales of Three Cities &  1884 &      M &                  no &          no \\
1727    &              Russell &                        Marooned: A Novel &  1889 &      M &                  no &         yes \\
7447    &               Brontë &                      Agnes Grey: A Novel &  1847 &      W &                  no &          no \\
8031    &                Grant &      Frank Hilton: or, "The Queen's Own" &  1855 &      M &                  no &          no \\
12062   &                Craik &        Cousin Trix and her Welcome Tales &  1868 &      W &                  no &          no \\
2551    &             Mackenna &                       A Child of Fortune &  1875 &      M &                  no &         yes \\
14966   &                Hearn &                   The Cathedral's Shadow &  1871 &      W &                  no &          no \\
1278    &              Edwards &                       Pharisees: A Novel &  1884 &      W &                  no &         yes \\
5683    &               Spicer &                          Bound to Please &  1867 &      M &                  no &         yes \\
8532    &           Boldrewood &    A Romance of Canvas Town and Other... &  1898 &      M &                  no &          no \\
12957   &              Shipton &                         Bearing the Yoke &  1884 &      W &                  no &          no \\
11917   &             Clifford &                Studies in Brown Humanity &  1898 &      M &                  no &          no \\
7430    &            Ainsworth &  The Star-Chamber: An Historical Romance &  1854 &      M &                  no &         yes \\
3949    &              Chapman &                            Mary Bertrand &  1860 &      W &                  no &         yes \\
2231    &                Hardy &                         Friend and Lover &  1880 &      W &                  no &         yes \\
4217    &               Pardoe &                            Reginald Lyle &  1854 &      W &                  no &         yes \\
5152    &            Anonymous &    A Heart Well Won: or, The Life and... &  1874 &      U &                 yes &         yes \\
539     &                 Parr &                              Dorothy Fox &  1870 &      W &                  no &         yes \\
4026    &            Anonymous &   The Prince of Orange: A Tale of the... &  1845 &      U &                 yes &         yes \\
2664    &                Reade &          A Simpleton: A Story of the Day &  1873 &      M &                  no &         yes \\
7437    &                Yates &                   Two By Tricks: A Novel &  1874 &      M &                  no &         yes \\
9550    &             Phillips &  The Birth of a Soul: A Psychological... &  1894 &      W &                  no &          no \\
8371    &             Fletcher &                               Pasquinado &  1898 &      M &                  no &          no \\
\end{longtable}
\normalsize

\section*{Appendix: Novel definitions}
\label{appendix:novel-definitions}

Two definitions of the novel are used in large bibliographies of the 19th-century English novel:

\begin{itemize}
  \item ``RFGS'': a descriptive definition
found in \textcite{raven2000english} and \textcite{garside2000english}, and
  \item ``ATCL'': a more permissive definition than RFGS which includes all prose fiction of at least 90-printed pages that is not addressed exclusively to children.
\end{itemize}

RFGS adopts ``a more rigorous definition'' of the novel than previous scholars. Their
bibliography ``includes what contemporaries thought of as novels, incorporating works
categorized as `novels' in contemporary periodical reviews and under `novels' headings in
circulating library catalogues, but excludes religious tracts, chapbooks, literature
written only for children and juveniles, and very short separately issued tales.'' They
add the qualification: ``Collections of tales (including some mixed genre compilations)
are included; separate verse novels are not'' \autocite[4]{raven2000english}.

\textcite{garside2004english} also uses this definition of the novel.

ATCL uses a more inclusive definition of the novel. ATCL includes any prose fiction of at
least 90-printed pages but excludes any work aimed exclusively at children (younger than
12 years of age).  Hence, religious tracts or literature written for juveniles (12 years
of age or older) is included if it consists of prose fiction and meets the minimum length
requirements. These criteria are applied with considerable consistency. We believe that
exceptions (e.g., an 89-page novel), if any do appear in the database, occur in fewer than
1 in 200 records.

The works of author Eliza Paget offer an illustrative example in the application of these
two definitions.  Paget wrote ten prose fiction works between 1829 and 1841, all of which
are religious in theme and aimed at juvenile and adult readers.  Of those published before
1836, the bibliographies using the RFGS definition exclude all seven titles, categorizing
them as juvenile fiction.  Of those published after 1837, ATCL includes all three titles
since they are all prose fiction and meet the minimum length requirements.

\section*{Appendix: Probability-proportional-to-size group sizes}

\begin{longtable}{lrrr}
\toprule
  {} &  Man-authored & Woman-authored & Unknown \\
year &                   &                     &              \\
\midrule
1837 &        49 (0.2\%) &          32 (0.1\%) &  14 (0.06\%) \\
1838 &        49 (0.2\%) &          33 (0.1\%) &  14 (0.06\%) \\
1839 &        52 (0.2\%) &          36 (0.2\%) &  15 (0.07\%) \\
1840 &        51 (0.2\%) &          36 (0.2\%) &  15 (0.07\%) \\
1841 &        52 (0.2\%) &          38 (0.2\%) &  16 (0.07\%) \\
1842 &        54 (0.2\%) &          40 (0.2\%) &  17 (0.08\%) \\
1843 &        54 (0.2\%) &          42 (0.2\%) &  17 (0.08\%) \\
1844 &        56 (0.3\%) &          44 (0.2\%) &  18 (0.08\%) \\
1845 &        56 (0.3\%) &          45 (0.2\%) &  19 (0.09\%) \\
1846 &        59 (0.3\%) &          49 (0.2\%) &  20 (0.09\%) \\
1847 &        66 (0.3\%) &          55 (0.2\%) &   23 (0.1\%) \\
1848 &        69 (0.3\%) &          59 (0.3\%) &   25 (0.1\%) \\
1849 &        71 (0.3\%) &          62 (0.3\%) &   26 (0.1\%) \\
1850 &        77 (0.3\%) &          70 (0.3\%) &   29 (0.1\%) \\
1851 &        81 (0.4\%) &          74 (0.3\%) &   31 (0.1\%) \\
1852 &        82 (0.4\%) &          76 (0.3\%) &   32 (0.1\%) \\
1853 &        83 (0.4\%) &          80 (0.4\%) &   33 (0.1\%) \\
1854 &        87 (0.4\%) &          84 (0.4\%) &   35 (0.2\%) \\
1855 &        85 (0.4\%) &          84 (0.4\%) &   34 (0.2\%) \\
1856 &        82 (0.4\%) &          82 (0.4\%) &   33 (0.1\%) \\
1857 &        84 (0.4\%) &          85 (0.4\%) &   34 (0.2\%) \\
1858 &        84 (0.4\%) &          86 (0.4\%) &   34 (0.2\%) \\
1859 &        87 (0.4\%) &          91 (0.4\%) &   35 (0.2\%) \\
1860 &        90 (0.4\%) &          95 (0.4\%) &   37 (0.2\%) \\
1861 &        91 (0.4\%) &          95 (0.4\%) &   35 (0.2\%) \\
1862 &        92 (0.4\%) &          96 (0.4\%) &   35 (0.2\%) \\
1863 &        93 (0.4\%) &          98 (0.4\%) &   34 (0.2\%) \\
1864 &        98 (0.4\%) &         102 (0.5\%) &   35 (0.2\%) \\
1865 &       105 (0.5\%) &         113 (0.5\%) &   38 (0.2\%) \\
1866 &       105 (0.5\%) &         111 (0.5\%) &   35 (0.2\%) \\
1867 &       108 (0.5\%) &         114 (0.5\%) &   34 (0.2\%) \\
1868 &       114 (0.5\%) &         121 (0.5\%) &   35 (0.2\%) \\
1869 &       115 (0.5\%) &         123 (0.6\%) &   34 (0.2\%) \\
1870 &       118 (0.5\%) &         122 (0.5\%) &   31 (0.1\%) \\
1871 &       124 (0.6\%) &         136 (0.6\%) &   34 (0.2\%) \\
1872 &       124 (0.6\%) &         138 (0.6\%) &   33 (0.1\%) \\
1873 &       130 (0.6\%) &         147 (0.7\%) &   33 (0.1\%) \\
1874 &       127 (0.6\%) &        146.5 (0.7\%) &   32 (0.1\%) \\
1875 &       133 (0.6\%) &         162 (0.7\%) &   33 (0.1\%) \\
1876 &       142 (0.6\%) &         168 (0.8\%) &   33 (0.1\%) \\
1877 &       150 (0.7\%) &         177 (0.8\%) &   34 (0.2\%) \\
1878 &       158 (0.7\%) &         186 (0.8\%) &   35 (0.2\%) \\
1879 &       171 (0.8\%) &         202 (0.9\%) &   36 (0.2\%) \\
1880 &       175 (0.8\%) &         203 (0.9\%) &   37 (0.2\%) \\
1881 &       180 (0.8\%) &         204 (0.9\%) &   35 (0.2\%) \\
1882 &       182 (0.8\%) &         204 (0.9\%) &   34 (0.2\%) \\
1883 &       207 (0.9\%) &         226 (1.0\%) &   38 (0.2\%) \\
1884 &       220 (1.0\%) &         231 (1.0\%) &   38 (0.2\%) \\
1885 &       209 (0.9\%) &         218 (1.0\%) &   34 (0.2\%) \\
1886 &       219 (1.0\%) &         215 (1.0\%) &   36 (0.2\%) \\
1887 &       240 (1.1\%) &         226 (1.0\%) &   38 (0.2\%) \\
1888 &       269 (1.2\%) &         244 (1.1\%) &   42 (0.2\%) \\
1889 &       269 (1.2\%) &         236 (1.1\%) &   41 (0.2\%) \\
1890 &       290 (1.3\%) &         239 (1.1\%) &   41 (0.2\%) \\
1891 &       281 (1.3\%) &         229 (1.0\%) &   41 (0.2\%) \\
1892 &       311 (1.4\%) &         247 (1.1\%) &   45 (0.2\%) \\
1893 &       328 (1.5\%) &         251 (1.1\%) &   48 (0.2\%) \\
1894 &       343 (1.5\%) &         256 (1.1\%) &   50 (0.2\%) \\
1895 &       373 (1.7\%) &         277 (1.2\%) &   55 (0.2\%) \\
1896 &       370 (1.7\%) &         263 (1.2\%) &   55 (0.2\%) \\
1897 &       418 (1.9\%) &         292 (1.3\%) &   63 (0.3\%) \\
1898 &       422 (1.9\%) &         289 (1.3\%) &   64 (0.3\%) \\
1899 &       438 (2.0\%) &         293 (1.3\%) &   67 (0.3\%) \\
1900 &       434 (1.9\%) &         283 (1.3\%) &   69 (0.3\%) \\
1901 &       423 (1.9\%) &         275 (1.2\%) &   67 (0.3\%) \\
\hline
Total &    10,559 (47\%) &       9,436.5 (42\%) & 2293 (10\%) \\
\bottomrule
\caption{Counts of new novels published by year and author gender used in the
probability-proportional-to-size sampling strategy.  We use the medians of
the estimated counts in \textcite{riddell2018reassembling}. Counts shown here use the RFGS
definition of the novel, not the ATCL definition of the novel.}
\label{tbl:gender-year-cluster-sizes}
\end{longtable}

\end{document}